\documentclass[11pt]{article}

\usepackage{amsmath}
\usepackage{amssymb}
\usepackage{extarrows}
\usepackage{graphicx}
 \usepackage{indentfirst}
\usepackage{hyperref}

\renewcommand{\baselinestretch}{1.1}
  \renewcommand{\arraystretch}{1.0}
\begin{document}

 \title{A Note  on ``Confidentiality-Preserving Image Search:
A Comparative Study Between Homomorphic
Encryption and Distance-Preserving
Randomization"}

 \author{Zhengjun Cao$^1$ and  Lihua Liu$^2$}

     \footnotetext{$^1$Department of Mathematics, Shanghai University, Shanghai,
  China.  \\
     $^2$Department of Mathematics, Shanghai Maritime University,
  China.
   \textsf{liulh@shmtu.edu.cn}
    }

 \date{}\maketitle

\begin{quotation}
 \textbf{Abstract}. Recently, Lu et al. have proposed two image search schemes based on additive homomorphic encryption [IEEE Access, 2 (2014), 125-141].  We remark that both two schemes are flawed because:  (1) the first scheme does not make use of the additive homomorphic property at all;  (2) the additive homomorphic encryption in the second scheme is unnecessary and can be replaced by a more efficient symmetric key encryption.

 \end{quotation}

\section{Introduction}

Recently, Lu et al. \cite{L14} have discussed how existing
additive homomorphic encryption can be potentially
used for image search, and proposed two confidentiality-preserving image search schemes based on Paillier's encryption \cite{P99}.
In the proposed model, a client has many images who wants to store the image data online for convenient data access anywhere anytime. The client
 has to encrypt each image and its features and upload the encrypted data to a cloud server.
In this note, we remark that the Lu et al.'s schemes are flawed.
\section{Review of the Lu et al.'s schemes}

In the schemes \cite{L14}, the features of each image are encrypted by any additively homomorphic encryption such as Paillier's cryptosystem \cite{P99}, which can be described as follows.  Pick an RSA modulus $n=pq$. Set $\lambda = \mbox{lcm} (p-1, q-1)$.
 Select $g\in \mathbb{Z}^*_{n^2}$ such that  $n\,|\,\mbox{ord}_{n^2}(g)$.
  Publish $n, g$ and keep $\lambda$ in secret.   For $m\in \mathbb{Z}_{n}$, pick $r\in \mathbb{Z}_{n}$,  compute the ciphertext
  $c=\mathcal{E}(m)=g^mr^n \,\mbox{mod}\,  n^2.$  Recover $m=\mathcal{D}(c)=\left(\frac{c^{\lambda}-1 \,\mbox{mod}\,  n^2}{n}\right)/\left(\frac{g^{\lambda}-1 \,\mbox{mod}\,  n^2}{n}\right) \,\mbox{mod}\,  n $.

  Denote the encrypting  function and decrypting function of AES by $E(\cdot)$ and $D(\cdot)$, and that of Paillier's cryptosystem  by $\mathcal{E}(\cdot)$ and $\mathcal{D}(\cdot)$, respectively.  See Table 1 and Table  2  for the details of the two image search schemes.

\begin{center}
\renewcommand{\baselinestretch}{1.0}
  \renewcommand{\arraystretch}{1.0}
 \small \normalsize  \parskip 0mm

Table 1: The Lu et al.'s scheme 1  \vspace*{3mm}

 \begin{tabular}{|lcl|}
   \hline
   \fbox{Client} & & \fbox{Server} \\ \hline
 Encrypt the image $P^{(i)}$ and its  && \\
 feature vector $\textbf{f}^{(i)}\in \mathbb{R}^t$ as & & \\
 $\mathcal{E}(\textbf{f}^{(i)})=(\mathcal{E}(f^{(i)}_1), \cdots, \mathcal{E}(f^{(i)}_t))$  && Store the encrypted\\
and $E(P^{(i)})$. Upload them. &$ \xrightarrow[i=1, \cdots, N]{\{i, \mathcal{E}(\textbf{f}^{(i)}), E(P^{(i)})\}}$& images and features. \\
 \hline
Given an image $ Q$ and  its && \\
 feature vector $\textbf{q}$, ask for&& \\
  all the encrypted features. &$\xrightarrow{Request}$& \\
  &$\xleftarrow[i=1, \cdots, N]{\{i, \mathcal{E}(\textbf{f}^{(i)})\}}$ &  Return all encrypted features.  \\
 Compute  $\textbf{f}^{(i)}= \mathcal{D}(\mathcal{E}(\textbf{f}^{(i)}))$ and && \\
  the $L_2$ distance  $d_i=\parallel\textbf{f}^{(i)}-\textbf{q} \parallel$, && \\
   $i=1, \cdots, N$. Send && \\
  $\mathcal{I}=\{j\,|\, d_j\leq \lambda \}$, where $\lambda $ is && \\
  a fault-tolerant parameter. &$\xrightarrow{\  \mathcal{I} \ }$&  \\
 Recover all  & $\xleftarrow[k\in \mathcal{I}]{\{E(P^{(k)})\}}$ & Return all $E(P^{(k)}), k\in \mathcal{I}.$   \\
  $P^{(k)}=D(E(P^{(k)})), k\in \mathcal{I}.$   && \\  \hline
 \end{tabular}\end{center}

\begin{center}
\renewcommand{\baselinestretch}{1.0}
  \renewcommand{\arraystretch}{1.0}
  \small  \normalsize \parskip 0mm

Table 2: The Lu et al.'s scheme 2 \vspace*{3mm}

 \begin{tabular}{|lcl|}
   \hline
   \fbox{Client} & & \fbox{Server} \\ \hline
 Encrypt the image $P^{(i)}$ and its  && \\
 feature vector $\textbf{f}^{(i)}\in \mathbb{R}^t$ as & & \\
 $\mathcal{E}(\textbf{f}^{(i)})=(\mathcal{E}(f^{(i)}_1), \cdots, \mathcal{E}(f^{(i)}_t))$  && \\
and $E(P^{(i)})$. Compute && Store the encrypted\\
$\chi_i=\mathcal{E}\left(\Sigma_{\ell=1}^{t}(f^{(i)}_{\ell})^2\right) $.  Upload them.
 &$ \xrightarrow[i=1, \cdots, N]{\{i, \chi_i, \mathcal{E}(\textbf{f}^{(i)}), E(P^{(i)})\}}$& images and features. \\
 \hline
Given an image $ Q$ and   && Compute \\
its feature vector $\textbf{q}$, && $h_i= \left(\prod_{\ell=1}^{t} (\mathcal{E}(f^{(i)}_{\ell}))^{q_{\ell}}\right)^{-2}  $   \\
  send  $\textbf{q}$ to the server. &$\xrightarrow{\ \textbf{q}\ }$&  $\cdot\mathcal{E}\left(\Sigma_{\ell=1}^{t}q_{\ell}^2\right) \cdot\chi_i $, \\
  &&  $i=1, \cdots, N. $ \\
   Compute  $d_i= \mathcal{D}(h_i)$,   &$\xleftarrow{h_i, i=1, \cdots, N}$& Send them back. \\
       $i=1, \cdots, N$. Randomly      && \\
     pick a set $\widehat{\mathcal{I}}\subset \{1, \cdots, N\}$  && \\
     of an appropriate size. && \\
   Set $\mathcal{I}'=\widehat{\mathcal{I}}\bigcup \mathcal{I}$ where && \\
 $\mathcal{I}=\{j\,|\, d_j\leq \lambda, 1\leq j\leq N \}$, && \\
 $\lambda $ is a fault-tolerant parameter.
  &$\xrightarrow{\  \mathcal{I}' \ }$& \\
 Recover all images & $\xleftarrow[k\in \mathcal{I}']{\{E(P^{(k)})\}}$ & Return all   $E(P^{(k)}), k\in \mathcal{I}'.$   \\
  $P^{(k)}=D(E(P^{(k)})), k\in \mathcal{I}.$   && \\  \hline
 \end{tabular}\end{center}

Notice that, by the additive homomorphic property of Paillier's encryption, we have
\begin{eqnarray*}
 \mathcal{E}\left(\|\textbf{f}^{(i)}-\textbf{q} \|^2\right)&=& \mathcal{E}\left(\Sigma_{\ell=1}^{t}(f^{(i)}_{\ell}-q_{\ell})^2\right)
= \mathcal{E}\left(\Sigma_{\ell=1}^{t}(f^{(i)}_{\ell})^2-2\Sigma_{\ell=1}^{t}f^{(i)}_{\ell}q_{\ell}+\Sigma_{\ell=1}^{t}q_{\ell}^2\right)\\
  &=& \mathcal{E}\left(\Sigma_{\ell=1}^{t}(f^{(i)}_{\ell})^2\right)\cdot \mathcal{E}\left(-2\Sigma_{\ell=1}^{t}f^{(i)}_{\ell}q_{\ell}\right) \cdot \mathcal{E}\left(\Sigma_{\ell=1}^{t}q_{\ell}^2\right)   \\
   &=& \chi_i\cdot\left(\prod_{\ell=1}^{t} (\mathcal{E}(f^{(i)}_{\ell}))^{q_{\ell}}\right)^{-2} \cdot  \mathcal{E}\left(\Sigma_{\ell=1}^{t}q_{\ell}^2\right)=h_i,\\
  d_i&=& \mathcal{D}(h_i)= \mathcal{D}\left(\mathcal{E}( \|\textbf{f}^{(i)}-\textbf{q} \|^2)\right)=\|\textbf{f}^{(i)}-\textbf{q} \|^2.
  \end{eqnarray*}

\section{Analysis of the Lu et al.'s schemes}

We now show that the Lu et al.'s schemes are flawed.

\begin{itemize}
\item[(1)] \emph{The authors \cite{L14} have confused the general arithmetic over the field $\mathbb{R}$ and the modular arithmetic over the domain $\mathbb{Z}_{n}$}. In fact, the correctness of the schemes are based on
$$\textbf{f}^{(i)}= \mathcal{D}(\mathcal{E}(\textbf{f}^{(i)})), \ \ \|\textbf{f}^{(i)}-\textbf{q} \|^2=\mathcal{D}\left(\mathcal{E}( \|\textbf{f}^{(i)}-\textbf{q} \|^2)\right).$$
The equations hold on the condition that $\textbf{f}^{(i)}$ and $\|\textbf{f}^{(i)}-\textbf{q} \|^2$ are in the underlying domain $\mathbb{Z}_{n}$ of Paillier's encryption. That means a visual feature vector  $\textbf{f}\in \mathbb{R}^t$ must be transformed into $\widetilde{\textbf{f}}\in \mathbb{Z}_{n}^t$. But the authors \cite{L14} have not specified this process.

By the way, there is a typo in the description of the decrypting equation of Paillier's cryptosystem (see Ref.\cite{L14}, page 128). It should be
$m=\left(\frac{c^{\lambda}-1 \,\mbox{mod}\,  n^2}{n}\right)/\left(\frac{g^{\lambda}-1 \,\mbox{mod}\,  n^2}{n}\right) \,\mbox{mod}\,  n, $
not
$m=\left(\frac{c^{\lambda}-1 \,\mbox{mod}\,  n^2}{n}\right)/\left(\frac{g^{\lambda}-1 \,\mbox{mod}\,  n^2}{n}\right) \,\mbox{mod}\,  n^2 $.

\item[(2)] \emph{In the scheme 1, both the client and the server do not make use of  the additive homomorphic property of Paillier's encryption at all}. The  related computations for  the client are  $$\textbf{f}^{(i)}= \mathcal{D}(\mathcal{E}(\textbf{f}^{(i)})), \ i=1, \cdots, N.$$
  Actually, the process has no relation to the additive homomorphic property. Thus, in the scheme the Paillier's public key encryption can be reasonably replaced by the more efficient symmetric key encryption  AES.

  It seems that the authors \cite{L14} have not realized that the computational performance of public-key encryption is inferior to that of
symmetric-key encryption.  For example, the authors  wrote \cite{L14} ``image encryption can be done using state-of-the-art ciphers
such as AES or RSA by treating images as ordinary data".  We here would like to stress that images should be encrypted by a symmetric key encryption, instead of any public key encryption. In practice, RSA is usually  used for encrypting session keys, not for images. Compared with AES, RSA is fairly inefficient.

   %(see the Table 4 for the revised version of the scheme 1).
%
%  \begin{center}
%\renewcommand{\baselinestretch}{1.0}
%  \renewcommand{\arraystretch}{1.0}
% \small \normalsize  \parskip 0mm
%
%Table 4: The revised version of Lu et al.'s scheme 1  \vspace*{3mm}
%
% \begin{tabular}{|lcl|}
%   \hline
%   \fbox{Client} & & \fbox{Server} \\ \hline
% Encrypt the image $P^{(i)}$ and its  && \\
% feature vector $\textbf{f}^{(i)}\in \mathbb{Z}_{n}^t$ as & & \\
% $E(\textbf{f}^{(i)})=(E(f^{(i)}_1), \cdots, E(f^{(i)}_t))$  && Store the encrypted\\
%and $E(P^{(i)})$. Upload them. &$ \xrightarrow[i=1, \cdots, N]{\{i, E(\textbf{f}^{(i)}), E(P^{(i)})\}}$& images and features. \\
% \hline
%Given an image $ Q$ and  its && \\
% feature vector $\textbf{q}$, ask for&& \\
%  all the encrypted features. && Upon the request, send \\
%  &$\xleftarrow[i=1, \cdots, N]{\{i, E(\textbf{f}^{(i)})\}}$ &   all encrypted features back. \\
% Compute  $\textbf{f}^{(i)}= D(E(\textbf{f}^{(i)}))$ and && \\
%  the $L_2$ distance  $d_i=\parallel\textbf{f}^{(i)}-\textbf{q} \parallel$ && \\
%  for all $i=1, \cdots, N$. Send && \\
%  $\mathcal{I}=\{j\,|\, d_j\leq \lambda \}$, where $\lambda $ is && \\
%  a fault-tolerant parameter. &$\xrightarrow{\  \mathcal{I} \ }$& Return all the encrypted  \\
% Recover all images  & $\xleftarrow[k\in \mathcal{I}]{E(P^{(k)})}$ & images $E(P^{(k)}), k\in \mathcal{I}.$ \\
%  $P^{(k)}=D(E(P^{(k)})), k\in \mathcal{I}.$   && \\  \hline
% \end{tabular}\end{center}

 \item[(3)] In the scheme 2, the server has to make use of the additive homomorphic property for computing the encrypted distance $h_i=\mathcal{E}\left(\|\textbf{f}^{(i)}-\textbf{q} \|^2\right)$. But in such case, the client has still to compute $d_i= \mathcal{D}(h_i), i=1,\cdots, N$, which dominate the client's computational cost.  Compared with the revised scheme, we find,  \emph{the scheme 2 has not truly mitigated the client's computational cost}.  See the Table 3 for the comparisons of the dominated computations for the client in the three schemes.) Apparently, the revised scheme is more efficient because it only needs to perform symmetric key  decryption $N+|\mathcal{I}|$ times.

      \begin{center}
\renewcommand{\baselinestretch}{1.0}
  \renewcommand{\arraystretch}{1.0}
 \small \normalsize  \parskip 0mm

Table 3: The dominated computations for the client in the three schemes  \vspace*{3mm}

     \begin{tabular}{|l|l|l|}
   \hline
    & Dominated computations & Computational cost  \\ \hline
  Scheme 1 &  $\textbf{f}^{(i)}= \left(\mathcal{D}(\mathcal{E}(f_1^{(i)})), \cdots, \mathcal{D}(\mathcal{E}(f_t^{(i)}))\right),$
  &   public key  decryption: $tN$ (times)  \\
 &\quad  $i=1, \cdots, N$. &\\
 &  $P^{(k)}=D(E(P^{(k)})), k\in \mathcal{I}.$ & symmetric key  decryption: $|\mathcal{I}|$   \\ \hline
   Scheme 2 & $d_i= \mathcal{D}(h_i)$, $i=1, \cdots, N$.  & public key  decryption: $N$ \\
  &  $P^{(k)}=D(E(P^{(k)})), k\in \mathcal{I}.$ & symmetric key  decryption: $|\mathcal{I}|$
 \\ \hline
  The revised &  $\textbf{f}^{(i)}= D(E(\textbf{f}^{(i)}))$, $i=1, \cdots, N$. &  \\
  &  $P^{(k)}=D(E(P^{(k)})), k\in \mathcal{I}.$  &  symmetric key  decryption: $N+|\mathcal{I}|$ \\  \hline
 \end{tabular}
  \end{center}

\end{itemize}
\section{Conclusion}
We show that the Lu et al.'s schemes for image search are flawed and somewhat misleading. We here want to stress that
the computational performance of public-key encryption is inferior to that of
symmetric-key encryption. A homomorphic encryption allows anyone to perform some computations on encrypted data, despite not having the secret decryption key. But any computations performed on encrypted data are constrained to the underlying domain (finite domains). The real goal of using modular arithmetic in cryptography is to obscure and  dissipate the redundancies in a plaintext message, not to perform any numerical calculations.


\begin{thebibliography}{4}

\renewcommand{\baselinestretch}{1.0}
  \renewcommand{\arraystretch}{.9}
  \normalsize \small \parskip 1mm

\bibitem{L14} W.J. Lu, A. L. Varna and M. Wu, ``Confidentiality-Preserving Image Search:
A Comparative Study Between Homomorphic Encryption and Distance-Preserving Randomization", IEEE Access, 2 (2014), 125-141.


\bibitem{P99} P. Paillier, ``Public-Key Cryptosystems Based on Composite Degree Residuosity
Classes",  In: Stern, J. (ed.), Proc. of EUROCRYPT 1999, LNCS, vol. 1592, pp. 223-238, 1999.




\end{thebibliography}
\end{document}